# Response to "Comment on 'Ultradense protium p(0) and deuterium D(0) and their relation to ordinary Rydberg matter: a review'[Physica Scripta 94 (2019)075005]"

Leif Holmlid, email: holmlid@chem.gu.se

Department of Chemistry and Molecular Biology

University of Gothenburg, 412 96 Göteborg, Sweden

**Abstract**

In this answer to the Comment by Hansen and Engelen it is shown, that if there is any violation of the baryon number conservation law in H(0) nuclear reactions, it is not at all of the form that the authors believe. Their belief is disproved by cited well-known scientific results from other groups. It is further shown that quantum mechanics in H(0) molecules is different than these authors believe, not formulated in kinetic energy terms but defined by angular momentum quantization. Repetition of experiments is required, not pondering by non-specialists.

## 1. Introduction

The quantum material ultradense hydrogen H(0) has been studied experimentally with several methods since 2008 and there are now 65 publications in refereed scientific journals on H(0). Skepticism in some physics circles has existed but there have been no publications which have disproved any of the experimental or theoretical results. Despite this, non-scientific activities have been started like campaigns on the web, letters to my university and my department with allegations about lacking radiation protection, and letters to fifteen scientific journals to retract my published papers (none has been retracted). I have as a scientist survived several skepticism periods, first concerning Rydberg states at surfaces, then Rydberg matter, then H(0) and now baryon annihilation. After around 50 publications from me in each of these fields, the skepticism has faltered, but not yet for baryon annihilation due to its important relation to energy production.

I believe in scientific method and have always improved my work to counter the skepticism. I have often been told that my experiments are so simple that anyone can do them. Why are they then not repeated and published with their positive or negative outcome? Why do I not receive questions or other legitimate communications, if problems exist with the repetition, instead of letters to my university? I think the reason is just ignorance, but that opens for non-scientific activities like the ones mentioned above.

So now I answer this new comment by H+E as I will call them here to save space. I will answer scientifically and will not react further to their nice words like "egregious statements and inferences".

From their list of keywords it seems that their comment includes something on hydrogen phase diagrams but this is not the case. So, I answer with the same structure as H+E have used.

## 2. Baryon number conservation

"Besides the claim of 'cold fusion' of deuterons in ultra-dense deuterium…."

I do not claim cold fusion but muon-catalyzed fusion which is a well-known process since 1957. If other persons attribute my results to cold fusion it is outside my control. To see my view on cold fusion, please read Ref. 1.

"that of the conservation of the total number of protons and neutrons in the universe".

H+E must intend to write the sum of the number of protons and neutrons since free neutrons decay after 15 min to protons. The idea that the number of baryons is constant in the universe is sometimes given in popular science but it is not correct as every physicist should know. The law of baryon number conservation is something different. If in doubt, check with Wikipedia as a starter. I will mention a few pieces of evidence against the version of the law assumed by H+E.

First of all, a large mass of antimatter is assumed to have existed shortly after our Universe started to exist and each annihilation for example antiproton + proton removed one proton from existing. We do not know how much antimatter is still left in the Universe and how many protons are annihilated all the time. Even if we count antiprotons as baryons the total number of baryons is still not constant.

Secondly, particle accelerators can create baryons. For example antiprotons have been used in numerous experiments especially to study their annihilation reactions. Such studies both increase the number of baryons when the antiprotons (+ protons) are formed (or according to definition the baryon number is constant since antiprotons count negative) and then the number of baryons is decreased in the annihilation (or according to definition it is constant since +1-1 = 0).

So the baryon number conservation that H+E believe in does not exist. They state "A violation of this conservation law has never been observed.Not in high energy particle physics experiments, nor in another other type of experiment." The baryon number conservation law is that antiproton + proton counts +1-1= 0 thus the baryon number is conserved in a baryon annihilation, not as H+E believe that protons and antiprotons cannot disappear. Antiproton annihilation experiments can for example be found in Ref. 2.

"They have all shown no proton decay…."

Now H+E change the subject to proton decay which is a different process. I have never stated that the protons in our experiments decay. Why do they argue about protons?

On the other hand, the reaction equation they cite in my publication from 2019 is not correct. We had observed kaons at high intensity in the experiments and had to suggest some explanation for their formation. The truth was much more complex than I could know then. The complete process has been published "open access" in 2021 (Refs. 3 and 4) but H+E have apparently not read it. Their discussion about quarks is pointless and meaningless.

For example, they wonder how the strangeness in the kaons can start to exist. They apparently do not know that these particles are normally formed in pairs of one particle and its antiparticle. This of course avoids violation of any conservation laws and explains how the strangeness in the kaons appears. The main error in the reaction in my 2019 paper was that I proposed 3K to be formed. I should have proposed 2K (kaons) + 2$\pi$ (pions) but we had no final experimental evidence for the pions at that time.

"their suggested explanation completely disregards the difference between matter and antimatter"

This difference is often quite small. Neutral kaons are formed easily in the baryon annihilations and they are their own antiparticles thus there is no difference between matter and antimatter. The $D^0$ mesons oscillate according to CERN between matter and antimatter forms. So there is very little difference between matter and antimatter.

Another error in Eq. (1) is that the reaction was suggested to be p + p. As published in 2021 the correct reaction is p + anti-p. It took numerous measurements of meson energies before this conclusion was reached.

## 3. Molecular structure of $H_2$

Why this part of the comment is included is difficult to understand since the review from 2019 is not concerned with covalently bonded hydrogen molecules $H_2$. H+E do not mention any quantum numbers, molecular orbitals or angular momenta as important for this molecule but only its energy terms. Most of their discussion is not worth arguing about since a correct discussion can be found in many textbooks. I will just highlight a few of their statements:

"In their Fig.1 the authors provide their understanding of how this very short bond length can come about. Their argument is that the molecule has six Coulomb interactions….".

The reason we included Fig. 1 was that we received comments that H(0) could not exist due to the strong repulsions at short distance. Fig. 1 has nothing to do with the nature of H(0) otherwise which is completely given by angular momentum quantization. H+E apparently misunderstand this.

"Somehow there must be a repulsive force at work."

Of course, the repulsive terms are all included in the discussion of Fig. 1.

"By the argument of HZG, there is therefore nothing to prevent this atom from collapsing to a structure where the electron is located on top of the nuclear charge."

How this can be deduced from the review cannot be understood. The process H+E imply to be impossible indeed exists and is called beta capture. It seems that H+E think that the electron in an s orbital circles around the nucleus. They have probably not understood that the most likely location for an electron with $l = 0$ is in the nucleus. I have used this point as a test of student's understanding of quantum mechanics for a long time. The main factor which governs the electron motion in an atom is its quantized angular momentum which H+E does not mention at all. Maybe this is the reason why they do not accept H(0), which is defined by its quantized angular momenta as expressed very clearly in the review.

"With a total of four elementary particles in the molecule(two protons and two electrons), each of spin 1/2, it is impossible for any pairs of thosespins, and in particular the spins of the electrons, not to be aligned. This will render theirpostulated cancellation of the Coulomb interaction inoperative."

It is apparent that the authors do not believe in pairing of electrons with different spins, up and down as it is often called. They should consider the basic knowledge that an orbital has the same shape whether there are one or two electrons in it. Thus the cancellation which we observe in passing (it is not necessary for our argument) exists and is well known. There exist many good textbooks on quantum mechanics where the elementa of molecular theory can be found.

## 4. Experiments, measurements

"employing a standard, potasium doped, iron oxide catalyst"

The catalyst is a crucial part of the experiments and to lightly call it a standard catalyst shows great ignorance. It cannot be bought, so what makes it a standard catalyst?

"The postulated bond length of a few picometers,in particular, is calculated from the observation of flight times in mass spectra."

This sentence contains numerous errors.The H-H bond length is not postulated but measured. The calculation is based on the kinetic energy release from Coulomb explosions in <u>neutral</u> time-of-flight spectra, not in mass spectra. Please observe this point very carefully: Coulomb explosions of the molecules give the kinetic energies of the fragments. The time-of-flights of the neutral fragments are measured and this gives the kinetic energy of the fragments which is several hundred eV.

"No attempt has been documented of any attempt to rule out anyother explanation, for example the obvious suggestion that the spectra are due to chargingup of the sample."

Such data are indeed given in the publications. but are not observed by H+E. How charging up of a metallic grounded sample could give nanosecond time-of flight spectra requires new physics. The spectra are reproducible; otherwise they would not have been studied or published. The sample is pure metal. Changes in its applied voltage shift the TOF-MS spectra correctly. This proves conclusively that there is no charging effect in the ion mass spectra. No charging effect is of course possible in the <u>neutral</u> time-of flight spectra.

"The peakswhich, noted in passing, have extremely poor resolution,….."

The resolution is determined by the kinetic energy release in the H(0) molecules and cannot be influenced by experimental parameters. The resolution is certainly good enough for determining the bond distances with considerable precision and also to derive the molecular shapes.

"One is that this production rate corresponds to an energy outputof close to 240 kW with an input of 5 W laser light."

The measurement cited is correct but the value of 240 kW looks unfamiliar. The experiment has been repeated many times over a period of several years and published a few times.An energy gain of 1000 is normal, not 50 000 as H+E state.

"Another is that it is made withoutany reference to radiation protection measures that should have been taken. This type ofintensities will cause serious damage to living biological matter in the surroundings and evento the experimental equipment used."

These comments are unprofessional. Further H+E know nothing about the radiation protection in my lab. The radiation that leaves the apparatus from H(0) is mainly neutral kaons and muons which give little radiation damage. This has been published in several papers. I have had manuscripts rejected since I was not dead and could not be right about the reactions in H(0). However, instead of fantasizing I measured the particles and the radiation. So I am still alive contrary to expectations and I know why.

## 5. Final comments

"The paper of Holmlid and Zeiner-Gundersen makes claims that would be truly revolutionary ifthey were true."

The claims are true. I suggest that you try such experiments yourself. You have the resources and if you have problems I offer help. Please note that the review was based on close to 50 published papers. Why should you be able to spot all the errors that must have been made in measuring and publishing these 50 papers if they are not correct?

"We have shown that they violate some fundamental and very well establishedlaws in a rather direct manner."

H+E probably mean the baryon number conservation law which they misunderstand as shown above. No other fundamental law was referred to. Their discussions about the hydrogen molecule demonstrate no understanding of quantization of angular momentum in atomic and molecular physics. Such quantization is fundamental for the theory of H(0).

## Conclusions

The comment by H+E could have touched upon some aspect of the review paper that summarized the results from close to 50 published papers. H+E do not have any argument against the review as such.They mainly discuss points that have been published in several other papers. So how that can be the content of a comment on the review paper is difficult to understand.

It is notable that H+E do not remark on that part of the review which concerns the rotational spectroscopy measurements, where a new summary with a complete table of the results was included. These measurements show very clearly picometer distances in p(0), D(0) and pD(0) for spin quantum numbers s = 2, 3 and 4. The uncertainty in the distances is down to a few femtometers. The published bond distance in state s = 2 is 2.245 ± 0.003 pm. This proves beyond any doubt that H(0) exists and has pm sized interatomic distances. These results may of course be difficult to understand for physicists with no experience in molecular rotational spectroscopy.